\documentclass[11pt,a4paper]{article}
\usepackage{graphicx}
\begin{document}

\begin{center}
{\Large SKYRME-HFB CALCULATIONS IN COORDINATE SPACE FOR THE
        KRYPTON ISOTOPES UP TO THE TWO-NEUTRON DRIPLINE}

\bigskip \bigskip

\small

V.E. OBERACKER$^1$, A. BLAZKIEWICZ$^2$, A.S. UMAR$^3$

\bigskip

\footnotesize

{\em $^1$Department of Physics and Astronomy, Vanderbilt University,
     Nashville, Tennessee 37235, USA,
E-mail: volker.e.oberacker@vanderbilt.edu\\
$^2$Department of Physics and Astronomy, Vanderbilt University,
     Nashville, Tennessee 37235, USA,
E-mail: artur.r.blazkiewicz@vanderbilt.edu\\
$^3$Department of Physics and Astronomy, Vanderbilt University,
     Nashville, Tennessee 37235, USA,
E-mail: umar@compsci.cas.vanderbilt.edu}

\bigskip

\small

(Received \today ) 

\end{center}

\bigskip

\footnotesize

{\em Abstract.\/} For axially symmetric even-even nuclei, we solve
the Hartree-Fock-Bogoliubov (HFB) equations on a 2-D grid in
cylindrical coordinates. The Skyrme SLy4 interaction is
used for the mean field and a zero-range interaction for the pairing field.
After decoupling the HFB lattice equations, we obtain quasiparticle states with
equivalent single-particle energies up to 100 MeV or more. We present
results for the Krypton isotope chain up to the two-neutron dripline,
including two-neutron separation energies, pairing gaps, and quadrupole
deformations for the ground states and isomeric minima.

\bigskip

{\em Key words:\/} Krypton isotopes ($A=104-116$), two-neutron dripline, HFB.

\normalsize


\section{INTRODUCTION \label{int} }

Experimental studies of the limits of nuclear stability provide stringent
tests for nuclear structure theories. While nuclei near the proton
dripline have been well explored, little is known about the limits of
nuclear binding on the neutron-rich side and about the exact location
of the neutron dripline. Only for the eight lightest nuclei has the neutron
dripline been reached, and the nuclear chart exhibits thousands of nuclear
isotopes still to be examined with the next generation of Radioactive Ion
Beam facilities. Another limit to nuclear stability is the superheavy element
region which is formed by a delicate balance between strong Coulomb repulsion
and additional binding due to closed shells.

For very light nuclei with $A \leq 12$, ab initio calculations have been carried
out using nucleon-nucleon and three-nucleon interactions. For heavier nuclei,
approximations are required; the most widely used approach is the self-consistent
mean field method.  Both non-relativistic mean field theories
\cite{DN96,TH96,RD99,SD00,YM01,TOU03,DS04} and relativistic versions
\cite{Ri96,LR99,KB00} have been developed.

In the vicinity of the driplines, pairing correlations increase dramatically
and it is essential to treat both the mean field and the pairing field
self-consistently within the Hartree-Fock-Bogoliubov (HFB) formalism \cite{DN96}.
In this region, not only does one have to consider ``well-bound'' single-particle states
but also ``weakly-bound'' states with large spatial extent, giving rise to neutron skins. All of these features represent major challenges for the numerical solution.
While the HF(B) theories describe the ground state properties of nuclei,
their excited states can be obtained with the (quasiparticle) random phase
approximation (Q)RPA \cite{Ma01,BD02,IJ03}.

Traditionally, the HFB equations have been solved by expanding the quasiparticle
wavefunctions in a truncated harmonic oscillator basis \cite{ER93}. This works very well near
the line of $\beta$-stability because only ``well-bound'' states need
to be considered. However, as one approaches the driplines, the numerical
solution becomes more challenging: in practice, it is very difficult to represent continuum
states as superpositions of bound harmonic oscillator states because the former
show oscillatory behavior at large distances while the latter decay exponentially.
In this case, coordinate-space representations have a distinct advantage
because ``bound'' and ``continuum'' states (in the box) are treated on an
equal footing, with no bias towards bound states.


\section{HFB EQUATIONS IN COORDINATE SPACE}

Over a period of several years, our research group has developed the
HFB-2D-LATTICE code, written in Fortran 95, which solves the HFB equations for deformed,
axially symmetric even-even nuclei in coordinate space on a 2-D
lattice \cite{TOU02,TOU03,OUTB03,BOU05,BOUS05,B05}.
The novel feature of our HFB code is
that it generates high-energy continuum states with an equivalent
single-particle energy of more than 100 MeV. QRPA calculations
require the inclusion of continuum states with such high energy
for the description of collective giant multipole resonances.

A detailed description of our theoretical method has been published in
Ref.~\cite{TOU03}; in the following, we give a brief summary.
In coordinate space representation, the HFB Hamiltonian and the quasiparticle
wavefunctions depend on the distance vector ${\bf r}$, spin projection
$\sigma = \pm \frac{1}{2}$, and isospin projection $q = \pm \frac{1}{2}$
(corresponding to protons and neutrons, respectively).
In the HFB formalism, there are two types of quasiparticle wavefunctions,
$\phi_1$ and $\phi_2 $, which are bi-spinors of the form
\begin{equation}\label{bispinor}
\phi^q_{1,\alpha} ({\bf r}) =
\left(
\begin{array}{c}
\phi^q_{1,\alpha} ({\bf r,\uparrow})\\  
\phi^q_{1,\alpha} ({\bf r,\downarrow})
\end{array}
\right) , \ \ 
\phi^q_{2,\alpha} ({\bf r}) =
\left(
\begin{array}{c}
\phi^q_{2,\alpha} ({\bf r,\uparrow})\\  
\phi^q_{2,\alpha} ({\bf r,\downarrow})
\end{array}
\right) .
\label{ec1}
\end{equation}
Their dependence on the quasiparticle energy $E^q_\alpha$ is denoted by the index
$\alpha$ for simplicity.
The quasiparticle wavefunctions determine the normal density
\begin{equation}
\rho_q({\bf r}) = \sum_{E^q_\alpha > 0}^{\infty} \sum_{\sigma = -\frac{1}{2}}^{+\frac{1}{2}}
       \phi^q_{2,\alpha} ({\bf r} \sigma) \ \phi^{q \ *}_{2,\alpha} ({\bf r} \sigma)
\label{eq:density} 
\end{equation}
and the pairing density
\begin{equation}
\tilde{\rho_q}({\bf r}) = - \sum_{E^q_\alpha > 0}^{\infty} \sum_{\sigma = -\frac{1}{2}}^{+\frac{1}{2}}
       \phi^q_{2,\alpha} ({\bf r} \sigma) \ \phi^{q \ *}_{1,\alpha} ({\bf r} \sigma) \ . 
\label{eq:pairing_density} 
\end{equation}
In the present work, we use the Skyrme SLy4 effective N-N interaction \cite{CB98}
for the mean field (p-h and h-p channels), and a zero-range pairing force (p-p and h-h channels). The pairing strength has been adjusted to reproduce the measured average
neutron paring gap of $1.245$ MeV in $^{120}$Sn. 
For these types of effective interactions, the particle mean field Hamiltonian $h$ and the
pairing field Hamiltonian $\tilde h$ are diagonal in isospin space and
local in position space
\begin{equation}
h({\bf r} \sigma q, {\bf r}' \sigma' q') \ = \ \delta_{q,q'}
      \ \delta({\bf r} - {\bf r}') h^q_{\sigma, \sigma'}({\bf r}) 
\end{equation}
and
\begin{equation}
\tilde{h}({\bf r} \sigma q, {\bf r}' \sigma' q') \ = \ \delta_{q,q'}
      \ \delta({\bf r} - {\bf r}') \tilde{h}^q_{\sigma, \sigma'}({\bf r}) \ .
\end{equation}
In spin-space, the mean field Hamiltonian is represented by the $2 \times 2$ matrix
\begin{equation}
h^q ({\bf r}) =
\left( 
\begin{array}{cc}
h^q_{\uparrow \uparrow}({\bf r}) & h^q_{\uparrow \downarrow}({\bf r}) \\
h^q_{\downarrow \uparrow}({\bf r}) & h^q_{\downarrow \downarrow}({\bf r})
\end{array}
\right)
\end{equation}
which operates on the bi-spinor wavefunctions in eq.~(\ref{bispinor}).
The pairing field Hamiltonian $\tilde h$ has a similar mathematical structure.
The HFB equations have the following structure in spin-space \cite{TOU03}:
\begin{equation}\label{hfb_coupled}
\left( 
\begin{array}{cc}
( h^q -\lambda^q ) & \tilde h^q \\
\tilde h^q & - ( h^q -\lambda^q ) 
\end{array}
\right)
\left(
\begin{array}{c}
\phi^q_{1,\alpha} ({\bf r})\\  
\phi^q_{2,\alpha} ({\bf r})
\end{array}
\right)
 = E^q_\alpha
\left( 
\begin{array}{c}
\phi^q_{1,\alpha} ({\bf r})\\  
\phi^q_{2,\alpha} ({\bf r})
\end{array}
\right)
\label{eq:hfbeq2}
\end{equation}
where $\lambda^q$ denotes the Fermi level. From now on, we drop the isospin
label $q$, for simplicity. The HFB equations may be recast in the form
\begin{equation}
H \phi_{\alpha} = E_{\alpha} \phi_{\alpha}
\end{equation}
with the four-spinor wavefunctions
\begin{equation}\label{eq:fourspinor}
\phi_{\alpha} ({\bf r}) = \left( 
\begin{array}{c}
\phi_{1,\alpha} ({\bf r})\\  
\phi_{2,\alpha} ({\bf r})
\end{array}
\right) = \left( 
\begin{array}{c}
\phi_{1,\alpha} ({\bf r,\uparrow})\\  
\phi_{1,\alpha} ({\bf r,\downarrow})\\
\phi_{2,\alpha} ({\bf r,\uparrow})\\  
\phi_{2,\alpha} ({\bf r,\downarrow})
\end{array}
\right)
\end{equation}
and the quasiparticle ``super-Hamiltonian'' ($4 \times 4$ matrix in spin
space) is given by
\begin{equation}\label{Hsuper}
H = \left(
\begin{array}{cc}
(h - \lambda)  & \tilde h \\
\tilde h & -(h - \lambda)
\end{array}
\right) \ .
\end{equation}
This quasiparticle Hamiltonian has the well-known property \cite{RS80,BR86} 
that for every eigenvector $\phi_{\alpha}^{+} = (U_\alpha,V_\alpha)$
with positive eigenvalue $E_{\alpha}^{+}$, there is a corresponding
eigenvector $\phi_{\alpha}^{-} = (V_\alpha^*,U_\alpha^*)$ with negative eigenvalue
$E_{\alpha}^{-} = - E_{\alpha}^{+}$, respectively
\begin{equation}\label{HFB_phi_pm}
H \phi_{\alpha}^{\pm} = E_{\alpha}^{\pm} \phi_{\alpha}^{\pm} \ .
\end{equation}
It is forbidden to choose the $E_{\alpha}^{+}$ and $E_{\alpha}^{-}$ solutions
at the same time, otherwise it is impossible to fulfill the anti-commutation
relations for the quasiparticle operators \cite{RS80}.
The quasiparticle energy spectrum is discrete for $|E_{\alpha}|<-\lambda$
and continuous for $|E_{\alpha}|>-\lambda$. For even-even nuclei it is customary to 
solve the HFB equations for positive quasiparticle energies and consider all negative
energy states as occupied in the HFB ground state \cite{DN96}. 


\section{2-D LATTICE REPRESENTATION}

In the case of even-even nuclei with axially symmetric shapes, we represent
the HFB equations in cylindrical coordinates ($\phi,r,z$) and eliminate the
dependence on the coordinate $\phi$, resulting in a reduced 2-D problem.
For a given angular momentum projection quantum number $\Omega$, we solve the
HFB equations on a 2-D grid $(r_\alpha,z_\beta)$ where $\alpha = 1,...,m$
and $\beta = 1,...,n$. In practice, angular momentum projections
$\Omega = 1/2, 3/2, ..., 21/2$ are taken into account. Typically, in radial
($r$) direction, the lattice extends from $0 - 15$~fm, and in symmetry axis ($z$)
direction from $-15,...,+15$~fm, with a lattice spacing of about $0.8$~fm in the central region. For the lattice representation of the HFB Hamiltonian we use a hybrid
method in which derivative operators are constructed using the Galerkin method
\cite{BOUS05}; this amounts to a global error reduction. Local potentials are
represented by the Basis-Spline collocation method \cite{UW91,WO95,KO96} (local
error reduction).

On the lattice, the four-spinor wavefunction in coordinate space $\phi^{(4)}(r,z)$
becomes an array $\phi(N)$ of length $N = 4 \times m \times n$
and the HFB Hamiltonian is transformed into a $N \times N$ matrix:
\begin{equation}
\sum_{\nu=1}^N H_{\mu}^{\ \nu} \phi^{\Omega}_{\nu} =
            E^{\Omega}_{\mu} \phi^{\Omega}_{\mu} \ \ \ (\mu=1,...,N) \ .
\end{equation}
The diagonalization of the HFB Hamiltonian using LAPACK routines yields 
quasiparticle states with energies up to 100 MeV or more. In the calculations
of observables, all quasiparticles states with equivalent single particle
energy up to the cut-off $\varepsilon_{max}=60$ MeV are taken into account.

Production runs of our HFB code are carried out on an IBM-SP massively parallel
supercomputer and on a local LINUX cluster using OPENMP/MPI message passing.
Parallelization is possible for different angular
momentum states $\Omega$ and isospins (p/n).


\section{DECOUPLED HFB EQUATIONS}

The HFB equations (\ref{hfb_coupled}) are coupled, i.e. the HFB Hamiltonian
in eq.~(\ref{Hsuper}) mixes the quasiparticle states $\phi_{1,\alpha}$ and
$\phi_{2,\alpha}$. We generalize the recipe given in Ref.~\cite{HZ97} to
decouple the HFB equations which results in a smaller diagonalization problem.

This is accomplished by the unitary transformation
\begin{equation}{\label{Utransf}}
U = R Z\;, \ \ \ \ \ \ 
R=\frac{1}{\sqrt{2}}
\left(
\begin{array}{cc}
 1 &1 \\
 1 & -1
\end{array}
\right)\;,\ \ \ \ \ \ 
Z=\frac{1}{\sqrt{2}}
\left(
\begin{array}{cc}
 1 &1 \\
 -i & i
\end{array}
\right)\;. 
\end{equation}
First we apply the complex unitary transformation $Z$
to the quasiparticle Hamiltonian resulting in a transformed Hamiltonian
in which the pairing fields on the diagonal appear with opposite sign
\begin{equation}
H_Z = Z H Z^{-1} = \left(
\begin{array}{cc}
\tilde h  &  +i h'\\
-i h' & -\tilde h
\end{array}
\right) \ ,
\end{equation}
where we have introduced the abbreviation
\begin{equation}
 h' = h - \lambda \ .
\end{equation}
Next we apply the real transformation $R$ and obtain a new
Hamiltonian $H_U$ in block off-diagonal form
\begin{equation}
H_U = U H U^{-1} = \left(
\begin{array}{cc}
 0 &  -i h' + \tilde h \\
 +i h' + \tilde h  &  0
  \end{array}
  \right)
\end{equation}
with the transformed eigenstates
\begin{equation}
\chi_{\alpha}^{\pm} = U \phi_{\alpha}^{\pm}
\end{equation}
resulting in 
\begin{equation}\label{huchi}
H_U \chi_{\alpha}^{\pm} = E_{\alpha}^{\pm} \chi_{\alpha}^{\pm} \ .
\end{equation}
Operating from the left with the transformed Hamiltonian $H_U$ onto
eq.~(\ref{huchi}) we obtain
\begin{equation}\label{hu2chi}
H_U^2 \chi_{\alpha}^{'} = E_{\alpha}^2 \chi_{\alpha}^{'} \ .
\end{equation}
It turns out that the operator $H_U^2$ has the desired block-diagonal form
\begin{equation}\label{hu2}
H_U^2 = \left(
\begin{array}{cc}
(-i h' + \tilde h) (+i h' + \tilde h ) & 0 \\
0 &  (+i h' + \tilde h) (-i h' + \tilde h )
\end{array}
\right) \ .
\end{equation}
The new eigenstates $\chi_{\alpha}^{'}$ are, in general, linear combinations
of the negative and positive  states
$\chi_{\alpha}^{\pm}$ because we lost
information about the sign of the energy E$_{\alpha}$ by squaring the
Hamiltonian
\begin{equation}
\chi_{\alpha}^{'} = w^{+} \chi_{\alpha}^{+} + w^{-} \chi_{\alpha}^{-}
\end{equation}
where $w^+$ and $w^{-}$ are complex constants.

From the structure of the Hamiltonian $H_U^2$ in eq.~(\ref{hu2}) we see that the
eigenstates $\chi_{\alpha,1}'$ and $\chi_{\alpha,2}'$ in eq.~(\ref{hu2chi})
are now decoupled. Therefore, we only need to diagonalize 
the upper and lower blocks of the Hamiltonian $H_U^2$ separately, resulting in
a speed-up by nearly a factor of 4 in the numerical calculations
compared to the original coupled diagonalization problem, eq.~(\ref{hfb_coupled}).
Further details regarding the numerical solution of the decoupled HFB equations
can be found in Ref.~\cite{B05}.


\section{RESULTS FOR THE KRYPTON ISOTOPES}

Recently we have studied the ground state properties of the
neutron-rich Zirconium isotopes ($Z=40$) up to the two-neutron
dripline \cite{BOU05,BOUS05}. In particular, we have calculated binding energies,
two-neutron separation energies, normal densities and pairing
densities, mean square radii, quadrupole moments, and pairing gaps.
In Ref.~\cite{BOUS05}, two theoretical approaches were presented and
compared: a) our HFB-2D-LATTICE code, and b) an expansion in a
transformed harmonic oscillator basis with 20 shells (HFB-2D-THO)
by Stoitsov {\it et al.} \cite{SD00,DS04}. In general, both theoretical methods
were found to be in excellent agreement for the entire isotope chain.
In particular, $^{122}$Zr is predicted to be the dripline 
nucleus.

\begin{figure}[!htb]
\begin{center}
\includegraphics*[scale=0.45]{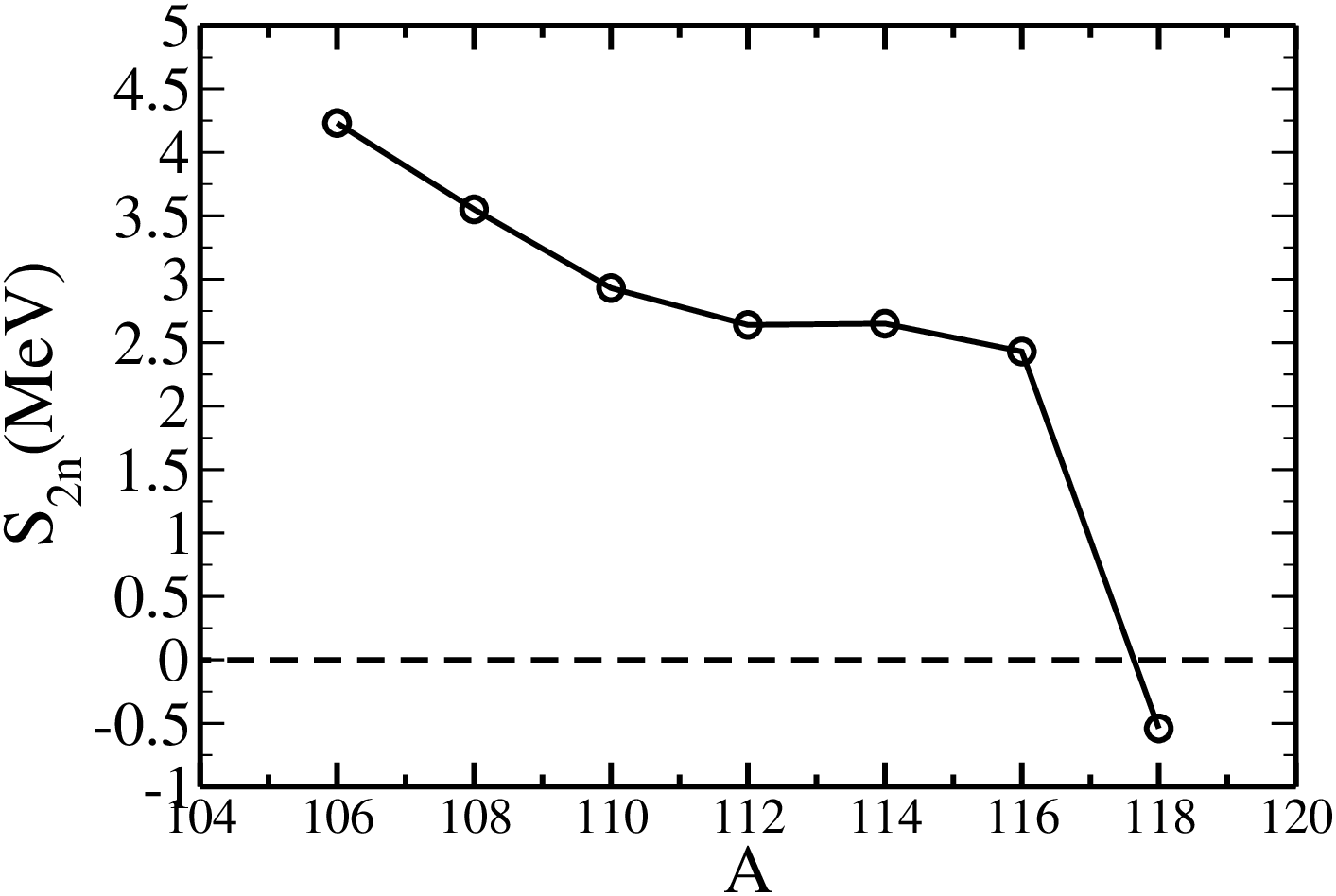}
\caption{\label{fig1} Two-neutron separation energies for
 the neutron-rich Krypton isotopes. The dripline is located where the separation
 energy becomes zero.}
\end{center}
\end{figure}

In this section, we present our HFB-2D-LATTICE code results for 
the ground state properties of the neutron-rich even-even Krypton isotopes
($Z=36$) in the mass region $A=104-118$. Fig.~\ref{fig1} shows the
two-neutron separation energies obtained from the calculated
binding energies. Experimental data for the Krypton isotopes calculated
in this work are not available because of the large N/Z ratios $\geq 1.8$.
Compared to the two-neutron separation energies for the Zr isotope chain
for the same mass number (see Fig.~1 of Ref.~\cite{BOUS05}) the separation
energy values are only about half the size. Our HFB calculations predict 
that $^{116}$Kr is the last stable nucleus against two-neutron emission.

Like the Zirconium isotopes studied in Ref.~\cite{BOUS05}, the Krypton isotope
chain reveals a rich variety of shapes. In particular, we observe
a transition from prolate to oblate shapes and back to prolate shapes 
in the mass region $A=104-118$. Specifically, the quadrupole deformations
decrease from $\beta_2$=0.38 ($^{104}$Kr) through  $\beta_2$=-0.31 
($^{108}$Kr) and then increase again until we reach
$\beta_2$=0.1 for $^{114,116}$Kr.
A similar trend for the shape transitions has been found 
by two other models, the Finite Range Droplet Model calculations
\cite{MoNI95}  and by the  Relativistic Mean-Field  calculations \cite{LR99}.
These latter two models use a simple BCS pairing interaction which is
not reliable for nuclei in the vicinity of the two-neutron dripline.

\begin{table}[!hbt]
\begin{center}
\caption{Shape coexistence in the Kr isotopes. The first column lists properties
of the ground state minimum (quadrupole deformation and binding energy), and
subsequent columns give results for shape isomeric states .\label{table:shapeKr}}
\vskip 0.25in
\begin{tabular}{||l  c|c|c| c|c| c| c|c|c| c|c|c|c|  l|| }
\hline
\hline
&  isotope   & \multicolumn{3}{c|}{1st min.}   & \multicolumn{3}{c|}{2nd min.}  &
         \multicolumn{3}{c|}{3rd min.}  \\
\cline{3-11}
&  & $\beta_{2}^{n}$ &$\beta_{2}^{p}$& E[MeV] & $\beta_{2}^{n}$ &$\beta_{2}^{p}$ &$E[MeV]$&
             $\beta_{2}^{n}$ &$\beta_{2}^{p}$& E[MeV]\\
\hline
\hline
& $^{104}$Kr &  0.38 & 0.35 &-821.41&  0.09  & 0.1 &-816.44& -0.26& -0.25&-821.30\\
\hline
& $^{106}$Kr &  0.36 & 0.35 &-825.65& - & - & - &-0.29& -0.28&-825.29    \\
\hline
& $^{110}$Kr & -0.15 &-0.15 & -832.14  &0.07 & 0.08&-830.45& 0.40& 0.36& -829.94  \\
\hline
\hline
\end{tabular}
\end{center}
\end{table}

We also find evidence for shape coexistence in several neutron-rich Krypton isotopes.
In Table (\ref{table:shapeKr}) we list the quadrupole deformation and binding energy
of the ground state PES minimum and several shape-isomeric minima
for the isotopes $^{104,106,110}$Kr. We notice that the energy difference between the  prolate and oblate minima for $^{104,106}$Kr are very small (of order of 10-40 keV).

A plot of the rms-radii for the Krypton isotope chain \cite{B05} shows the development
of a neutron skin as evidenced by the large difference between proton and
neutron rms-radii. A sudden rms-radius drop observed at A=108-110 coincides 
with the nuclear shape transition from prolate to oblate in this region.
A neutron skin development has also been seen in our HFB lattice calculations
of the Sulfur isotopes
(Fig.~4 in Ref.~\cite{OUTB03}) and in the Zr isotopes (Fig.~4 in Ref.~\cite{BOUS05}).

\begin{figure}[!htb]
\begin{center}
\includegraphics*[scale=0.4]{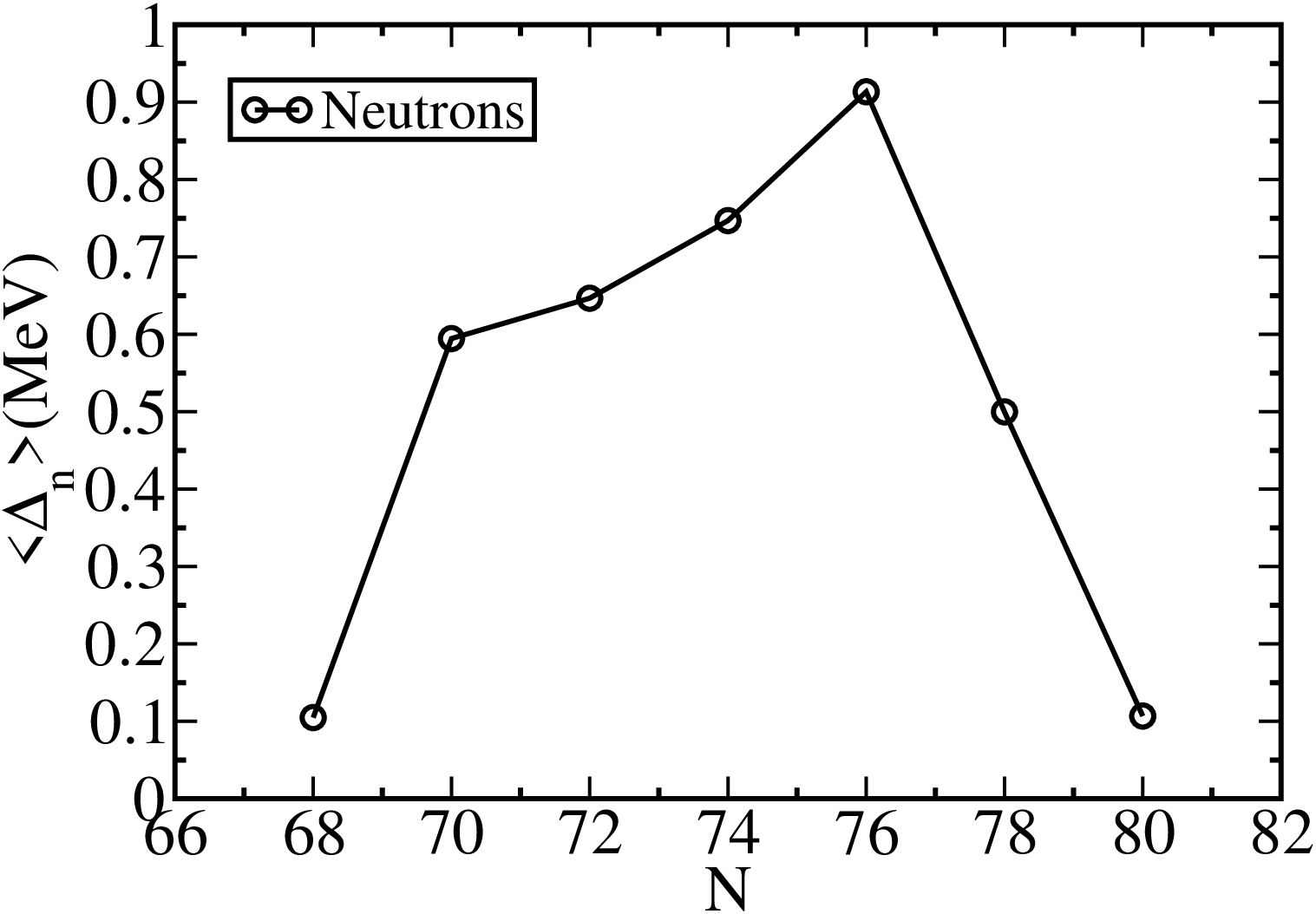}
\caption{\label{fig2}  Average neutron pairing gap
for the chain of Krypton  isotopes. }
\end{center}
\end{figure}

\begin{figure}[!htb]
\begin{center}
\includegraphics*[scale=0.4]{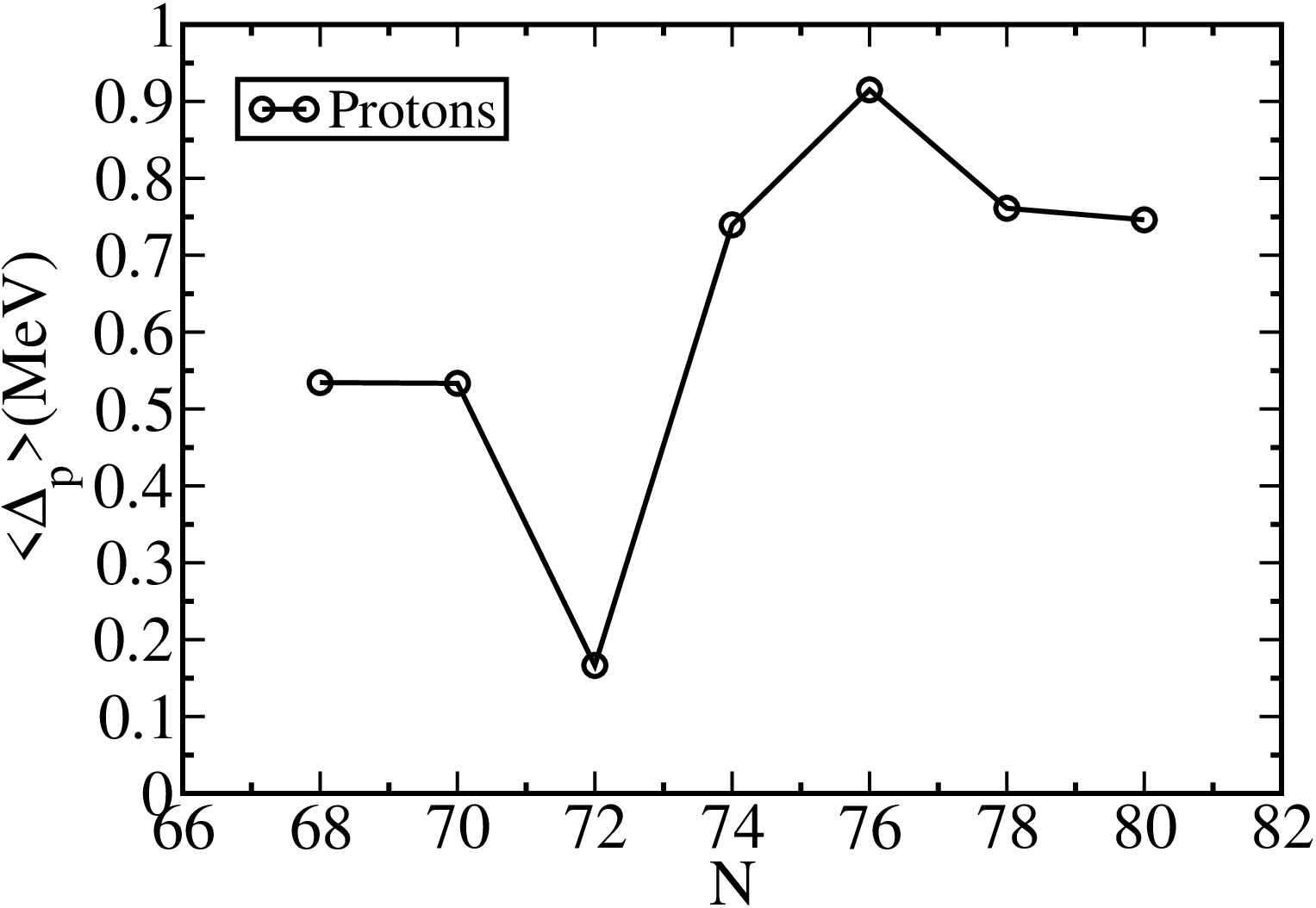}
\caption{\label{fig3}  Average proton  pairing gap
for the chain of Krypton  isotopes. }
\end{center}
\end{figure}

In general, the pairing densities and average pairing gaps for protons and
neutrons are the two observables which are most sensitive to the properties
of the continuum states. A study of the spectral distribution of the pairing
density for the Zirconium and Krypton isotopes \cite{OUBT04,B05} shows that
it is peaked at the Fermi energy $\lambda$  and reaches high into
the continuum for neutron-rich nuclei. In Figures \ref{fig2} and \ref{fig3}
we show the average pairing gaps for neutrons and protons, respectively.

We observe that the Krypton pairing gaps in the $A=104-118$ mass region vary
between 0.1 and 0.9 MeV. Starting at neutron number $N=68$, the neutron
pairing gap increases from 0.1 MeV until it reaches a maximum value of 0.9 MeV
at $N=76$; for higher neutron numbers it decreases quickly back to 0.1 MeV.
By contrast, the proton pairing gap shows an oscillatory structure in the
same mass region.


\section{REFERENCES}

\subsubsection*{Acknowledgements}
This work has been supported by the U.S. Department of Energy under grant
No. DE-FG02-96ER40963 with Vanderbilt University. Some of the numerical
calculations were carried out at the IBM-RS/6000 SP supercomputer of
the National Energy Research Scientific Computing Center which is
supported by the Office of Science of the U.S. Department of Energy.
Additional computations were performed at Vanderbilt University's ACCRE
multiprocessor platform.

\end{document}